\documentclass[manuscript]{acmart}

\AtBeginDocument{%
  \providecommand\BibTeX{{%
    \normalfont B\kern-0.5em{\scshape i\kern-0.25em b}\kern-0.8em\TeX}}}

\copyrightyear{2021} 
\acmYear{2021} 
\setcopyright{acmcopyright}\acmConference[IUI '21]{26th International Conference on Intelligent User Interfaces}{April 14--17, 2021}{College Station, TX, USA}
\acmBooktitle{26th International Conference on Intelligent User Interfaces (IUI '21), April 14--17, 2021, College Station, TX, USA}
\acmPrice{15.00}
\acmDOI{10.1145/3397481.3450645}
\acmISBN{978-1-4503-8017-1/21/04}

\begin{document}

\title{TeethTap: Recognizing Discrete Teeth Gestures Using Motion and Acoustic Sensing on an Earpiece}

\author{Wei Sun}
\authornote{Both authors contributed equally to the paper.}
\affiliation{%
    \institution{Institute of Software, Chinese Academy of Sciences}
    \country{China}
}
\affiliation{%
    \institution{Cornell University}
    \country{United States}
    }
\additionalaffiliation{%
  \institution{School of Computer Science and Technology, University of Chinese Academy of Sciences, China}
}

\author{Franklin Mingzhe Li}
\authornotemark[1]
\affiliation{%
    \institution{Carnegie Mellon University}
    \country{United States}
 }
 
\author{Benjamin Steeper}
\authornotemark[1]
\affiliation{%
    \institution{Cornell University}
    \country{United States}
 }
 
\author{Songlin Xu}
\affiliation{%
    \institution{Cornell University}
    \country{United States}
}
\affiliation{%
\institution{University of Science and Technology of China}
    \country{China}
}
\author{Feng Tian}
\affiliation{%
    \institution{School of Artificial Intelligence, University of Chinese Academy of Sciences, Beijing}
    \country{China}
}
\additionalaffiliation{%
    \institution{State Key Laboratory of Computer Science, Institute of Software Chinese Academy of Sciences, China}
}
 
\author{Cheng Zhang}
\authornote{Corresponding author}
    \affiliation{%
    \institution{Cornell University}
    \country{United States}
 }


\newcommand{\FL}[1]{\textcolor{red}{\textbf{FL:} #1}}
\newcommand{\WS}[1]{\textcolor{blue}{\textbf{Wei:} #1}}
\newcommand{\CZ[1]}{\textcolor{orange}{\textbf{Cheng:} #1}}

\newcommand{\TrackChange}[1]{#1}
\begin{abstract}
Teeth gestures become an alternative input modality for different situations and accessibility purposes. In this paper, we present TeethTap, a novel eyes-free and hands-free input technique, which can recognize up to 13 discrete teeth tapping gestures. TeethTap adopts a wearable 3D printed earpiece with an IMU sensor and a contact microphone behind both ears, which works in tandem to detect jaw movement and sound data, respectively. TeethTap uses a support vector machine to classify gestures from noise by fusing acoustic and motion data, and implements K-Nearest-Neighbor (KNN) with a Dynamic Time Warping (DTW) distance measurement using motion data for gesture classification. A user study with 11 participants demonstrated that TeethTap could recognize 13 gestures with a real-time classification accuracy of 90.9\% in a laboratory environment. We further uncovered the accuracy differences on different teeth gestures when having sensors on single vs. both sides. Moreover, we explored the activation gesture under real-world environments, including eating, speaking, walking and jumping. Based on our findings, we further discussed potential applications and practical challenges of integrating TeethTap into future devices.
\end{abstract}



\maketitle

\section{Introduction}

The vast-majority of input techniques for mobile devices demands the use of hands as an input source, which may constraints user experiences. For example, it would be inconvenient for a user to interact with a smartwatch to reject a phone call while both hands are occupied (e.g. carrying objects \cite{ng2013impact}). Therefore, providing hands-free interactions may improve the wearable interactive experiences under different situational uses and provide additional input opportunities for accessibility purposes (e.g., people with motor impairments).

To understand different hands-free interaction options, prior works explored eye-tracking systems \cite{Dhuliawala2016SmoothEM,Esteves2015OrbitsGI}, tongue input \cite{Saponas:2009:OST:1622176.1622209}, teeth input \cite{ashbrook2016bitey, mohamed2006teethclick}, facial expression \cite{masai2015affectivewear}, and voice recognition \cite{860214}. Eye-tracking systems \cite{Esteves2015OrbitsGI} usually require a mounted camera attached to glasses or a stationary device to become hands-free. Beyond having a camera facing users, more research leveraged head-worn sensors to track input gestures from the face. However, many of these works either require complex on-body sensor contacts on the face \cite{iravantchi2019interferi,rantanen2013capacitive,zhang2014non}, abnormal sensor locations \cite{mohamed2006teethclick} or placing sensors in the mouth \cite{gallego2019chewit,saponas2009optically}. To simplify the sensing requirements and hardware complexity, further research explored hands-free interaction techniques through sensors that are easily mounted on existing wearable devices, such as glasses or earbuds. For example, CanalSense leveraged barometers in the earbuds to classify face-related movements \cite{ando2017canalsense}. However, past works mostly explored face-related gestures through either barometers \cite{ando2017canalsense} or bone-conduction microphones \cite{ashbrook2016bitey}. Furthermore, many existing devices have already embedded inertial measurement unit (IMU) sensors, such as eSense \cite{kawsar2018esense}, into earpieces. However, acoustic-only approaches may have limitations related to noise from speaking, chewing or outside agents, and motion-only approaches may be affected while the user is in a motion (e.g., jogging). Furthermore, little research has explored the feasibility of leveraging motion sensing (e.g., IMU) combined with acoustic sensing (e.g., microphone) to recognize face-related gestures. 

\begin{figure}[t!]
\centering
\includegraphics[width=0.6\columnwidth]{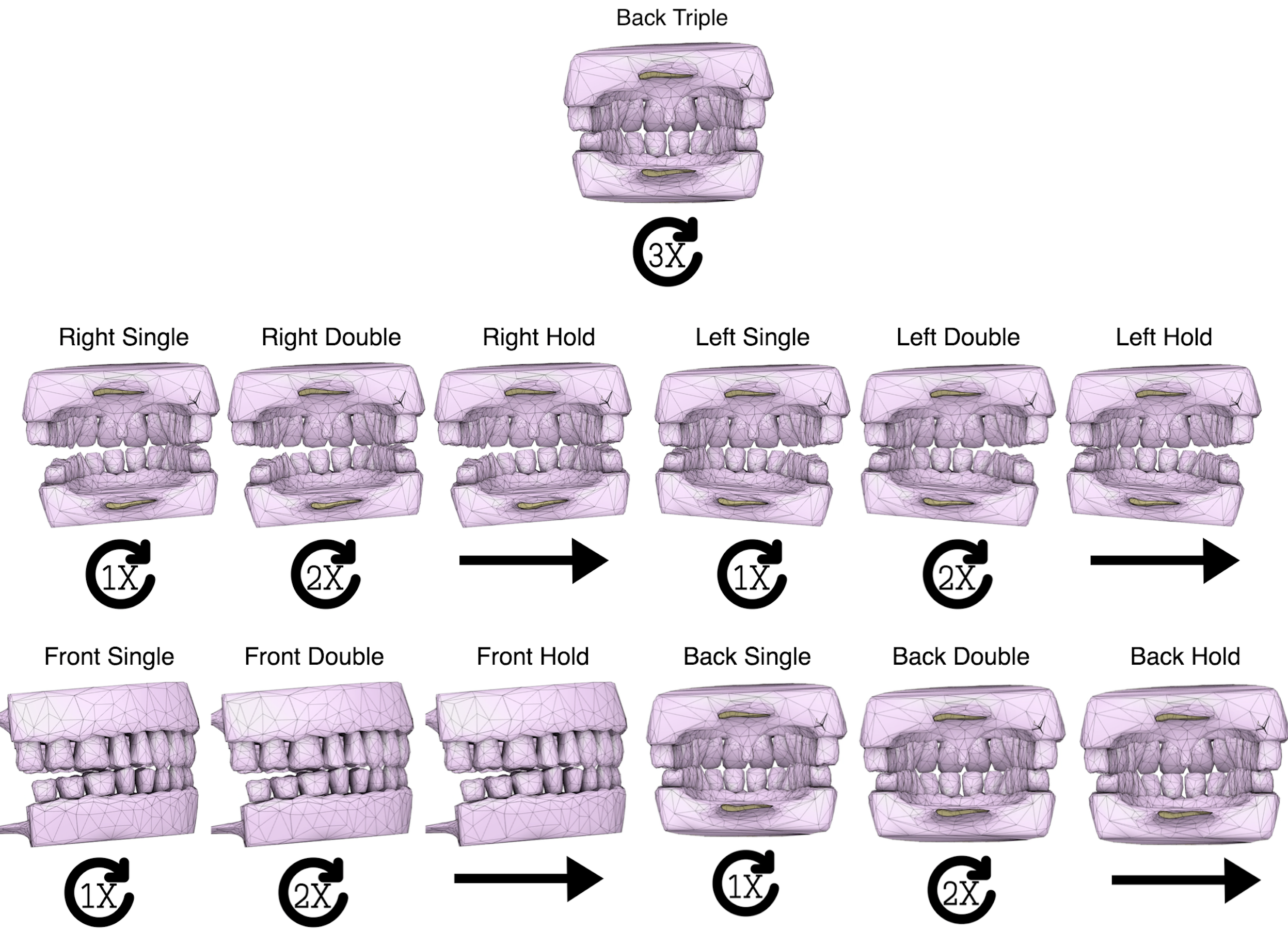}
\caption{Our gesture set with 13 teeth gestures.}
\label{fig:f1_hardware}
\Description{This figure contains thirteen teeth gestures with corresponding pictures in three rows. The first row: back triple. The second row: right single, right double, right hold, left single, left double, and left hold. The third row: front single, front double, front hold, back single, back double, and back hold.}
\end{figure}

In this paper, we present TeethTap, a minimally-obtrusive eyes-free, hands-free input technology that can recognize up to 13 discrete teeth gestures (Fig. \ref{fig:f1_hardware}), which cover both places of contact (i.e., left side, right side, front and back) and methods of contact (i.e., single bite, double bite, or hold). To recognize these 13 teeth gestures, we built a lightweight earpiece, which secures a microphone and IMU sensor behind each ear. The earpiece was made of 18 small components which were 3D printed and then fitted together, and was adjustable to various ear sizes and head widths. To understand the feasibility of leveraging TeethTap to recognize teeth gestures, we conducted a user study with 11 participants in five sessions (i.e., one practice session, one training session, two testing sessions, and one remounting session). We then analyzed the accuracy to recognize gestures using a DTW-based K-Nearest-Neighbor(KNN) algorithm, which has been widely used to classify IMU-based data in previous literature \cite{Zhang2017FingerSoundRU}.

Overall, TeethTap achieved 90.9\% accuracy on average to classify 13 discrete teeth input gestures in the testing sessions. We further compared the differences in the accuracy of having sensors on both sides vs. one single side. We found that it is sufficient to only use a single side sensor to recognize `manner' gestures, such as single-tap, double-tap, and hold. We also uncovered the accuracy differences in remounting the sensors by participants themselves and participants' subjective feedback. We further discussed the existing challenges of using TeethTap in-the-wild, the potential applications (e.g., volume control with ``Hold'' gestures), integrating TeethTap to other devices, and how to avoid remounting problems of TeethTap. We believe our findings shed light on future research that leverages motion and acoustic sensing on earpieces to recognize teeth gestures. Our contributions are summarized as follows:

\begin{itemize}
    \item We explored the feasibility of leveraging motion sensing captured around the ear, and fusing motion and acoustic signals to filter noises, to recognize 13 discrete teeth gestures with an average accuracy of 90.9\%.
    \item We uncovered the effect on different gestures of having motion sensors on both sides vs. one side and discussed the influences on recognition accuracy from remounting the devices.
    \item We proposed a set of design implications to apply the combination of motion and acoustic sensing on earpieces (e.g., in-the-wild scenario, design form factors, integrating to other head-worn devices).

\end{itemize}

\section{Related Work}

Hands-free interaction techniques benefit people under different scenarios. Prior research exploring hands-free wearable input devices focuses on tracking eye movement \cite{Barz2015ComputationalMA,Esteves2015OrbitsGI,Sugano2015SelfCalibratingHE,zhang2017smartphone,fan2020eyelid}, head movement \cite{Esteves2017SmoothMovesSP}, jaw movement \cite{taniguchi2018mouthwitch} and lip movement \cite{dalka2010human}. For example, Rantanen et al. \cite{rantanen2013capacitive} leveraged head-mounted capacitive and electromyography (EMG) sensors to detect different facial gestures. Similarly, Interferi \cite{iravantchi2019interferi} allowed users to wear a face sensing mask that used acoustic interferometry to track face-related gestures. 
However, these approaches often require heavy instrumentation on the user, such as cameras, magnets or headsets, to accurately distinguish between user input gestures. Recently, reserachers presented C-Face\cite{10.1145/3379337.3415879},  an ear-mounted wearable that can track facial movements, which has shown promising performance. But it is unclear how it can track teeth-input gestures. 

To explore other hands-free interaction techniques that require minimal hardware instrumentation and complexity, prior works explored different approaches to recognize teeth gestures \cite{ashbrook2016bitey} and tongue gestures \cite{nguyen2018tyth,Saponas:2009:OST:1622176.1622209,taniguchi2018earable}. Researchers first explored the approaches by adding sensors inside the mouth, such as embedded optical sensors into orthodontic dental retainers to detect tongue gestures \cite{saponas2009optically} and intraoral sensing bit to detect different tongue and teeth gestures \cite{gallego2019chewit}. Moreover, Li et al. \cite{Li:2013:STO:2493988.2494352} used sensor-embedded teeth to recognize four mouth-related activities: coughing, chewing, drinking and speaking. However, these approaches might be obtrusive to some people who do not have dental retainers or do not want to hold a sensor bit in the mouth.

To avoid placing sensors inside the mouth, past researchers further explored other approaches like placing bone-conduction microphones (e.g., \cite{ashbrook2016bitey}) on the skin to track teeth gestures or tongue gestures. For example, TeethClick \cite{mohamed2006teethclick} placed a single throat microphone that touched the cheek and picked up vibration signals from the jawbone to recognize single vs. double teeth clicks. To further make the hardware instrumentation less obtrusive, Bitey \cite{ashbrook2016bitey} recognized tooth click sounds from up to five different pairs of teeth gestures with bone-conduction microphones worn above the ears. However, Bitey tested user-specific gesture sets tailored to each participant, and the study relied solely on acoustic data, which has several limitations related to noise from speaking, chewing or outside agents. 

Another approach to track tooth-clicks is to use motion sensing (e.g., IMU). Simpson et al. \cite{4473368} introduced Tooth-Click Detector, which used a three-axis accelerometer on an earbud to pick up strong vibrations from tooth-clicks to control computer cursors. Zhao et al. \cite{Zhao:2012:TET:2168556.2168632} further employed the Tooth-Click Detector \cite{4473368} as well as an eye-gaze tracker to type on an on-screen keyboard. Researchers had also used tooth-touch sound as an alternative mouse device for accessibility \cite{Kuzume2011ToothtouchSA,Simpson2010EvaluationOT}. However, these approaches were binary---only being able to detect whether or not there was a tooth click. Recently, many existing earpieces have already embedded IMU sensors, such as eSense \cite{kawsar2018esense}, and have been applied for activity recognition \cite{katayama2019situation,lee2019automatic}. However, it is unknown whether it is feasible to detect different teeth gestures through earpieces with IMU sensors. 

Previous works also explored the combination of using IMU sensors and microphones to detect eating behaviors  \cite{Bedri:2017:EUW:3139486.3130902, Bi2018AuracleDE}. We understand that acoustic sensors are more error-prone to background acoustic noise, and motion sensors are more likely to have false positives while the user is in motion. Therefore, it is important to explore how both motion and acoustic sensors can be used in tandem on earpieces to detect different teeth gestures and reduce false positives. In our work, TeethTap fuses acoustic sensing with motion sensing to accurately classify a large set of 13 universally applied teeth gestures. By combining data from two separate sensing modalities into one device, our system is able to better realize noise and recognize teeth tapping movements. Furthermore, our instrumentation is minimally-intrusive, securing both sensors discreetly behind each ear. Strategic sensor placement combined with a robust classification system makes TeethTap a viable future accessory to the ear.

\section{Gesture Design}

Our approach in designing teeth gestures was inspired in part by two linguistic vowel sound features: the degree of aperture (jaw openness) and tongue frontness (or backness) \cite{prakash2020earsense}. The degree of aperture functions as z-axis, and is relevant for gesture release detection. We applied the idea of tongue frontness to the jaw, functioning as a y-axis. Lastly, we added a final axis (x-axis) for side-side movement. The four extremes of our x-y plane can therefore be described as front, back, left, and right. This design maximized the spread of each point of contact to best avoid confusion when classifying one gesture from another. 

In linguistics, there are two primary categories of articulation: the place of articulation and the manner of articulation \cite{gallego2019chewit}. As described above, our gestures have four places the teeth can make contact: front, back, left, and right. For each place of contact, TeethTap employs three possible ``manners'' of contact: single tap, double tap, and hold. ``Single tap'' is a quick tap and release. Naturally, ``double tap'' is composed of two quick single taps followed by a release. ``Hold'' is a tap with a delayed release. The time passed from the start of the hold gesture when the teeth first make contact, and the release of the gesture is registered as a continuous variable representing analog input. All non-hold gestures represent discrete digital inputs. We also added a gesture into our gesture set. This gesture is composed of three quick single back (regular) bites in sequence. We designed this gesture to be natural to produce yet easily recognizable for the purpose of testing under various real-world conditions such as walking, jumping, eating and speaking.

\section{Methodology}
\begin{figure}[h]
\centering
\includegraphics[width=0.15\columnwidth]{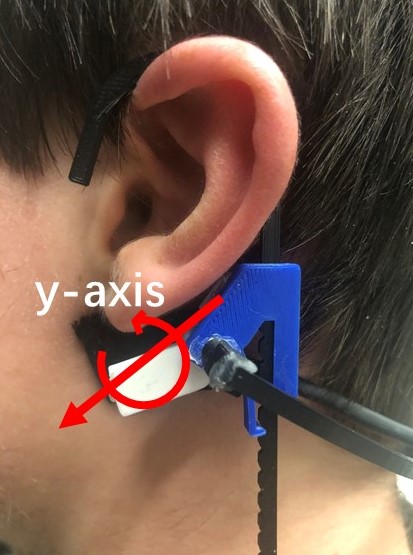}
\caption{Y-axis in relation to jaw movement}
\label{fig:JawMovement_yaxis}
\Description{This figure shows a person's ear with a 3D printed earpiece mounted. The figure also points out the direction of the y-axis.}
\end{figure}
\subsection{Sensing Principle}

By positioning our IMU sensors just behind the bottom of the ear where the jawline begins, we are able to collect gyroscope movement across three axes whenever the jaw shifts upwards, downwards or sideways. Fig. \ref{fig:JawMovement_yaxis} illustrates this principle on the IMU's y-axis under the left ear. As the jaw extrudes leftward, it presses against the bottom part of the left IMU, causing the gyroscope to rotate upwards. The resulting rotation causes its y-axis value to increase. On each earpiece, we also placed a microphone to collect and analyze acoustic data from different teeth gestures.

\begin{figure}[h]
\centering
\includegraphics[width=0.6\columnwidth]{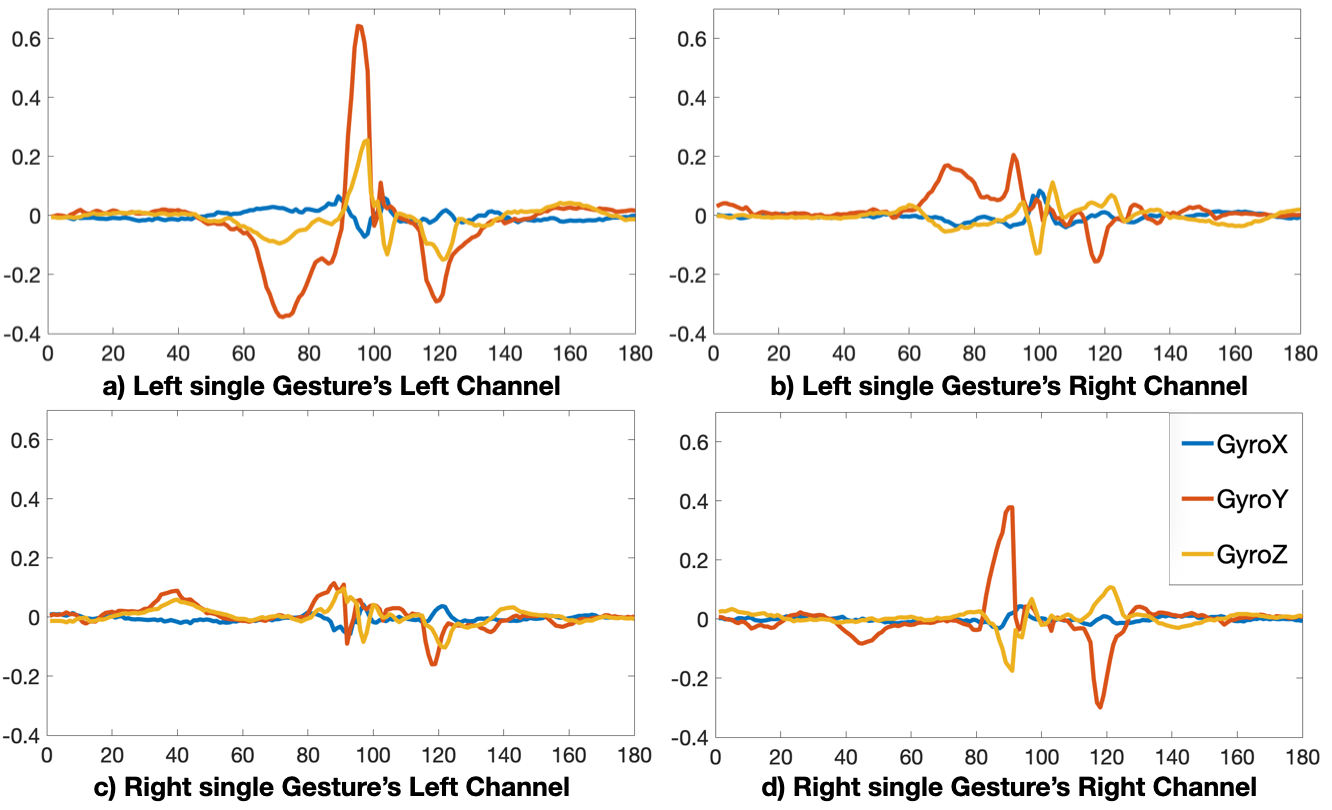}
\caption{Raw Gyroscope data for Left Single and Right Single Gestures}
\Description{This figure contains two by two sub-figures. The top left figure shows the left single gesture's signal from the left channel. The top right figure shows the left single gesture's signal from the right channel. The bottom left figure shows the right single gesture's signal from the left channel. The bottom right figure shows the right single gesture's signal from the right channel.}
\label{fig:GestureLeftRight}
\end{figure}

Fig. \ref{fig:GestureLeftRight}(a) shows the left ear's IMU data during a left single gesture, clearly depicting this positive y-axis peak. Conversely, the right ear IMU depicts a negative y-axis peak, as the right jaw retracts, rotating the right IMU in the opposite direction (Fig.  \ref{fig:GestureLeftRight}(b)). Similarly, Fig. \ref{fig:GestureLeftRight}(d) further shows a similar peak, this time illustrating the right ear's IMU data for a single right gesture. Again, the negative peak in Fig. \ref{fig:GestureLeftRight}(c) is caused by the left side of the jaw retracting and rotating the IMU downwards.

\begin{figure}[h]
\centering
\includegraphics[width=0.8\columnwidth]{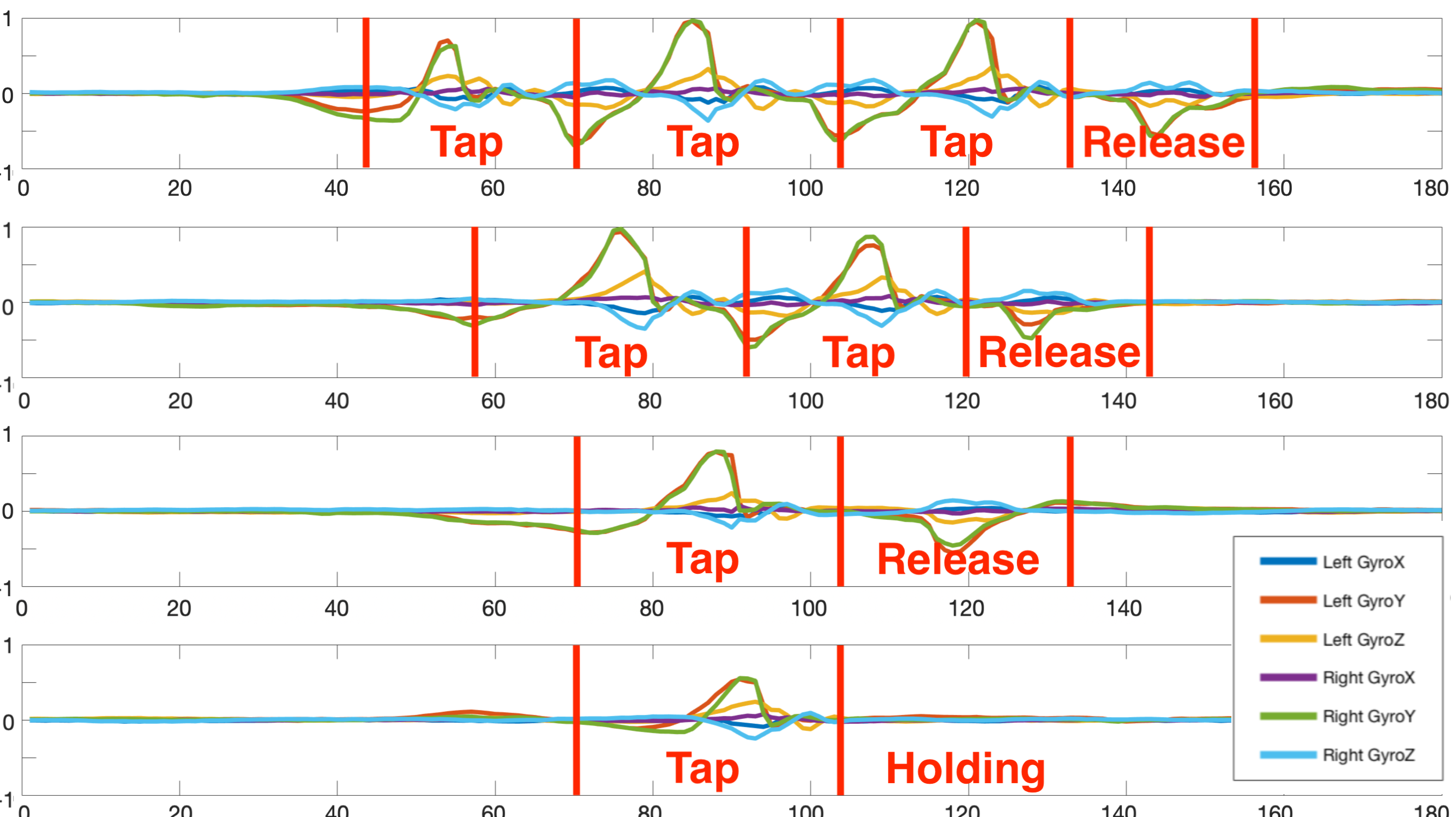}
\caption{Raw Gyroscope data for back triple, back double, back single, and back hold Gestures}
\label{fig:GestureComparison}
\Description{This figure contains four figures in a single column. The first figure at the top shows signals of tap, tap, tap, and release. The second figure shows signals of tap, tap, and release. The third figure shows signals of tap and release. The fourth figure shows signals of tap and holding.}
\end{figure}

Fig. \ref{fig:GestureComparison} illustrates the gyroscope data from four gestures: back triple, back double, back single, and back hold (top to bottom, respectively). The first three high-amplitude peaks in Figure \ref{fig:GestureComparison}(a) represent the back-triple gesture. The fourth smaller peak at the end of the window represents the gesture release. Release energy is captured when the mouth opens after performing a gesture. Back double (Fig. \ref{fig:GestureComparison}(b)) and back single (Fig. \ref{fig:GestureComparison}(c)) also end with a release peak. Notice that back hold (Fig. \ref{fig:GestureComparison}(d)), has no release peak because hold gestures delay the release, categorizing it instead as a separate sub-gesture.

\subsection{Hardware Design}

\begin{figure}[h]
\centering
\includegraphics[width=0.6\columnwidth]{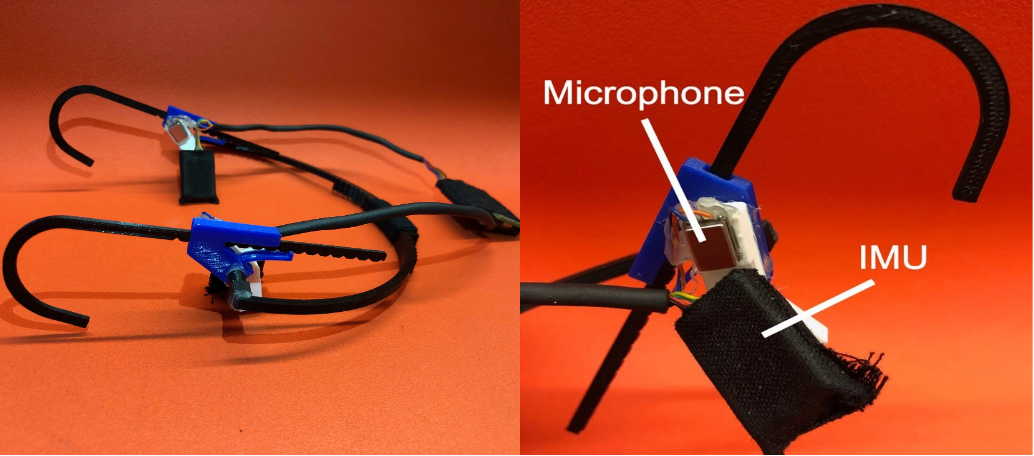}
\caption{3D printed earpiece housing two microphones and two IMUs}
\label{fig:earpiece1}
\Description{This figure contains two sub-figures. The left one shows the 3D printed earpiece from an overview. The right one shows the microphone and IMU of one side of the 3D printed earpiece.}
\end{figure}

TeethTap's hardware is composed of a 3D printed earpiece housing two contact microphones and two IMUs.  
Our 3D printed earpiece is made from 18 small individual components assembled together to form a single unit (Fig. \ref{fig:earpiece1}). The design is adjustable around the ears and behind the head to accommodate for various ear sizes and head widths. The natural flex of thinly printed PLA filament presses the IMU sensors against the jawline just under the ear and secures the microphones to the temporal bone behind the ear. We used two contact microphones (BU-30179-000) \cite{BU30179046:online} and two inertial measurement units (IMU) (MPU-9250) \cite{invensense2014mpu} to capture sound and motion on the skin behind each ear. The contact microphones are connected to a customized PCB board, which amplifies and filters the acoustic signals. The filtered data from acoustic sensors and the gyroscope data from IMUs are sent to a micro-controller (HUZZAH32) \cite{Adafruit58:online}
using its on-board 12-bit analog to digital converter (ADC) and its inter-integrated circuit (I$^2$C) communication, respectively. The microphone data is sampled at 8000 Hz, and the IMU data is sampled at 120 Hz. Lastly, the HUZZAH32 sends the data to a computer for processing using WiFi.

\begin{figure}[h]
\centering
\includegraphics[width=1\columnwidth]{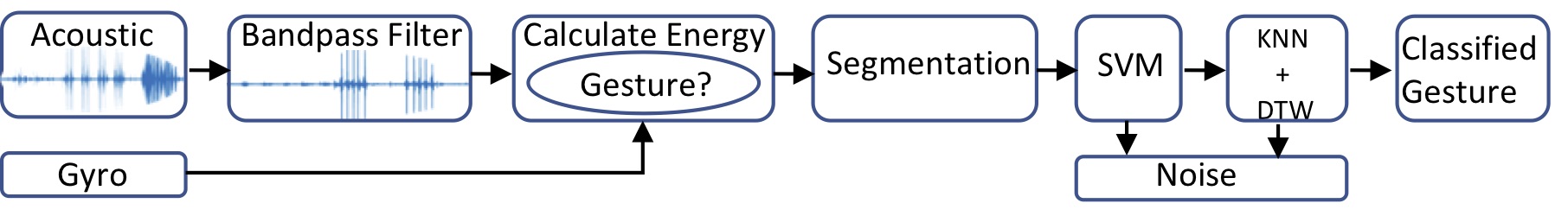}
\caption{Data Processing Pipeline}
\label{fig:dataprocessingpipleine}
\Description{This figure shows the data processing pipeline. From Acoustic to Bandpass Filter, to Calculate Energy and combined with the Gyro data to determine whether it is a gesture or not. Then it goes to segmentation, then SVM points to KNN + DTW, and both point to Noise. Finally, KNN + DTW points to Classified Gesture.}
\end{figure}

\subsection{Data Processing Pipeline}
To collect sensor data from the HUZZAH32 board, we created a Python program on the receiving computer.
We also used the same program to analyzes the data for gesture recognition in two stages: gesture segmentation and gesture classification. Figure \ref{fig:dataprocessingpipleine} illustrates TeethTap's data processing pipeline. 

\subsection{Gesture Segmentation}
Our algorithm first segmented a two-second sliding window from the continuous data stream generated from the microphones and IMUs. As data flowed in and out of the queue, our sliding window shifted 20 times a second with an overlap of 95 percent. Every window, we checked if the microphone data exceeded a predetermined energy threshold, which indicated a gesture was possibly performed. Once our system detected a sufficient spike in audio data, we then grabbed that window's corresponding two-second gyroscope data window. Next, we checked if the gyroscope's y-axis absolute maximum value exceeded a predetermined energy threshold to understand whether a gesture was performed. At this stage, we waited until the gesture was centred within the two-second sliding window in preparation for segmentation. Because most participants finished each gesture in roughly 1.5 seconds, further segmentation was needed. To segment the data, we smoothed the absolute value of the peak(s) to find the gesture's center-point and added a 90 data point buffer on each side to form a finalized event region of 1.5 seconds (i.e., 180 data points).

\subsection{Noise Detection with Acoustic Sensing}
Although TeethTap's contact microphone was hardly affected by outside noise, self-generated noise such as eating, talking or walking might interfere with the system. To address this issue, we implemented an SVM model classifier with a linear kernel to train both acoustic features and IMU features in the frequency domain. To collect acoustic data for noise and gestures, we asked one researcher and two pilot participants (one female) to each perform each teeth gesture five times, and we collected noise information by asking them to talk, walk, eat food, and remain static in random order. Overall, we collected 650 gesture segments and 650 noise segments. 

TeethTap extracted features from the IMU data and the microphone data for SVM classification. Seven of the eight IMU-related features were calculated across each of the three axes for both gyroscopes (six axes total). These included the number of peaks, peak values, root mean square (RMS), zero-crossing rate, standard deviation, minimum value, and maximum value. The eighth IMU-derived feature was calculated by finding the correlation between each of the left gyroscope axes with each of the right gyroscope axes. We also collected two acoustic features from the microphone data: the 30 lower bins of the Fast Fourier Transform (FFT) and 26 Mel-frequency cepstral coefficients (MFCC) \cite{Zhang2017FingerSoundRU}. This was made for a total of 64 features used to train our SVM model. We then applied the model to classify noise segments vs. gesture segments from acoustic data in TeethTap (Fig. \ref{fig:dataprocessingpipleine}).

\subsection{Gesture Classification Algorithm}
After segmenting the data and filtering out the noise, we classified the gestures by using K-Nearest-Neighbor (k=1) with a distance measurement of multi-dimensional Dynamic Time Warping (DTW) \cite{ten2007multi}. DTW is known for finding temporal patterns (similarities) between time-series datasets (especially with small training sets). Our first ran DTW (Dynamic Time Warping) on the data gathered during gesture segmentation with each gesture instance from training, one at a time. DTW's distance function would then output a value from every iteration. The gesture with the smallest distance value was determined as the predicted gesture.

\section{User Study}

\subsection{Participants}
\TrackChange{To evaluate the real-time gesture recognition performance and the usability of TeethTap, we issued a recruitment announcement on the school campus and recruited 11 participants with an average age of 24.3 (from 21 to 34, five female). Our participants are students or employees of the university, and all of them have healthy teeth.} Each participant received a $\$$10 gift card or cash for participating in our study. The study for each participant lasted around one hour and was conducted in a laboratory environment. The study was approved by the institutional review board (IRB).

\begin{figure}[h]
\centering
\includegraphics[width=0.7\columnwidth]{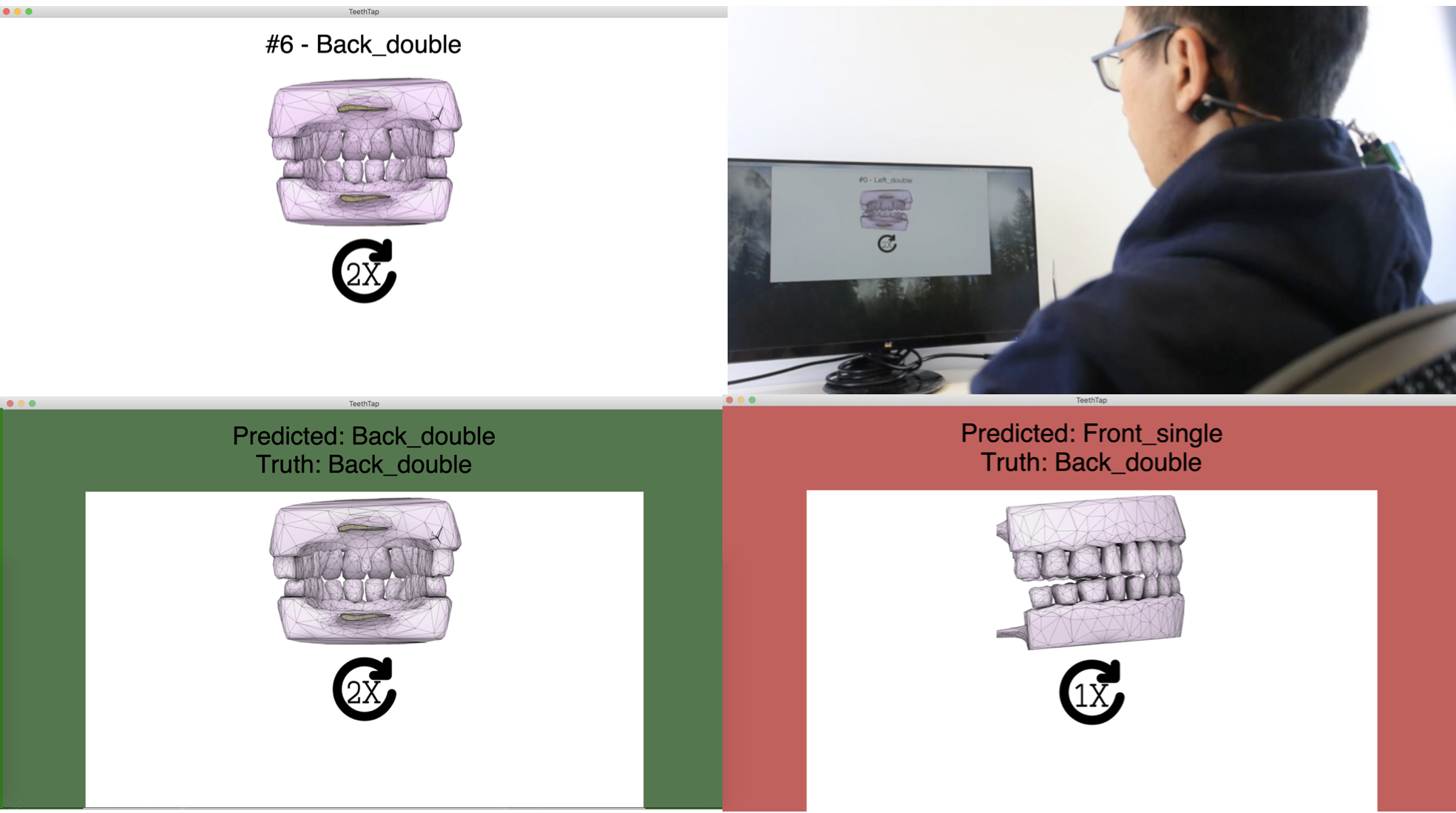}
\caption{The GUI of the user study}
\label{fig:GUI} 
\Description{This figure contains two by two sub-figures. The top left shows a teeth gesture of back double. The top right shows a person sit in front of the monitor. The bottom left shows a gesture with predicted and truth as back double with green background. The bottom right shows a gesture with predicted as front single and truth as back double. The background of the bottom right is red.}
\end{figure}

\subsection{Procedure}
At the beginning of the user study, we played a video that demonstrated how to perform 13 TeethTap gestures using animations of teeth constructed with AutoCAD (Fig. \ref{fig:f1_hardware}), followed by a live demonstration of the system by the researcher. Next, we helped the participant put on the device and explained the user interface (UI) of the system. The participant was asked to sit in front of a table with a monitor that displayed the testing UI. We then conducted the study in five different sessions: one practice session, one training session, two testing sessions, and one remounting session.

In each of the five sessions, participants were asked to perform each of the 13 gestures five times in a random order, which was indicated in the monitor. The first session was the practice session, which was designed to help the participant familiarize themselves with the gestures and testing system. The second session was the training session. The data collected in the training session was used to train an ML model, which was then used to provide real-time classification in later sessions. In the two testing sessions, we provided real-time classification results to the participant. The model was trained using the data collected in the training session. If the gesture was recognized as the same gesture appearing on the screen, we changed the background to green. Otherwise, the background was turned to red, and the recognized gesture's name and the picture were displayed. Whenever the system detected a holding gesture, the UI displayed a clock and asked the participants to hold for a randomly generated time interval (two to four seconds) until release. If no release gesture was detected within five seconds after the timer ended, the system timed out, counting the attempt as a recognition failure and proceeding to the next gesture in sequence.

To further understand the effect of taking TeethTap off and put it back on the same training data, we conducted a remounting session. Participants were asked to take off our prototype and put it back before this session started. Afterward, participants followed the same instructions as the testing session. In total, we collected 2860 (11*13*5*4) gesture instances in the training, testing and remounting sessions. At any point in the study, if the participant misconducted a gesture (performed a gesture than was different from the requested gesture), we asked the participant to report this to the researcher, and we removed these instances from the training and testing data. In total, 89 out of 2860 instances were removed. The real-time classification results and sensor data were saved for later analysis. After our participants finished the five sessions, we asked their subjective feedback on our system, potential applications, and improvements.

\subsection{Results}

\begin{figure}[h]
\centering
\includegraphics[width=0.7\columnwidth]{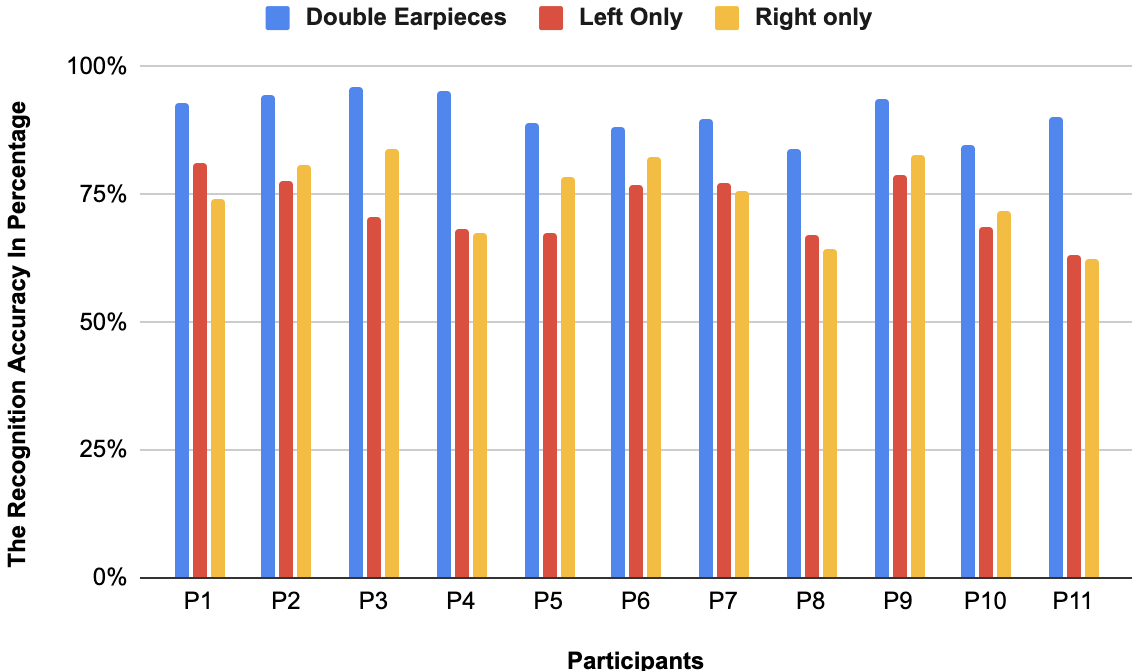}
\caption{The recognition accuracy for 11 participants in the testing sessions}
\label{fig:accuraciesFirsttwotest}
\Description{This figure is a bar chart of the recognition accuracy for 11 participants. X-axis: each participant. Y-axis: the recognition accuracy in percentage. For each participant, the figure shows three bars in blue (double earpieces), red (left only), and yellow (right only).}
\end{figure}

\subsubsection{Results from Both Earpieces}

\begin{figure}[h]
\centering
\includegraphics[width=0.8\columnwidth]{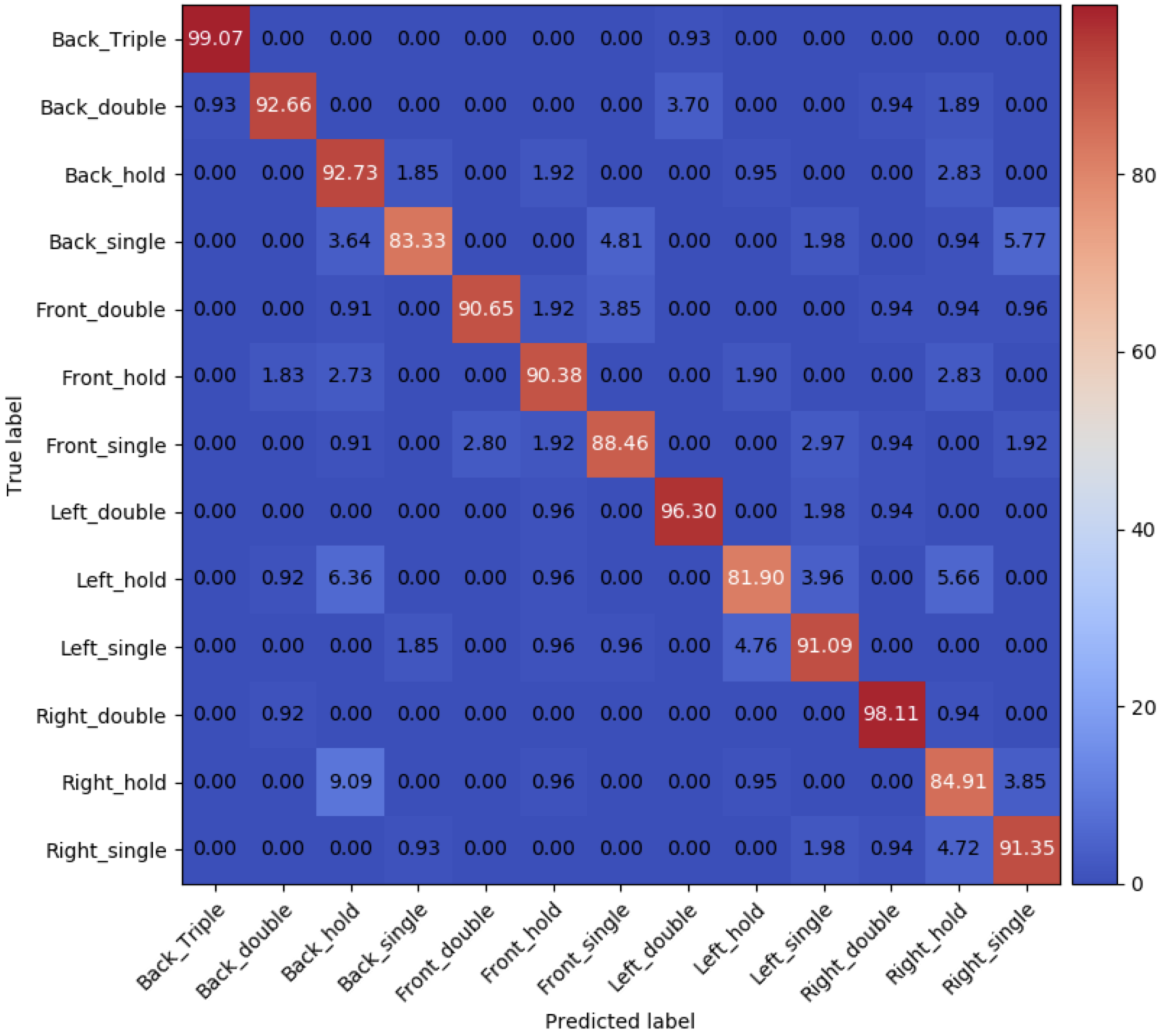}
\caption{The confusion matrix for recognition accuracy of the testing sessions}
\label{fig:ConfusionMatrix}
\Description{This figure shows a confusion matrix with rows are true labels and columns are predicted labels. For each value in the confusion matrix, the background color close to red if the value is close to 100 and the background color close to blue if the value is close to 0.}
\end{figure}

In the two testing sessions of recognizing 13 different gestures, we found that our participants reached an average accuracy of 90.9\% (SD = 4.1\%). Within the 1382 total teeth gestures from 11 participants, TeethTap successfully recognized 1256 gestures. Fig. \ref{fig:accuraciesFirsttwotest} demonstrates each participant's individual teeth gestures recognition accuracy. We found that P3 and P8 had the highest and lowest accuracy of 96.2\% and 83.9\%, respectively. There were only five holding gesture instances where the system failed to detect the release gesture. As shown by the confusion matrix presented in Fig. \ref{fig:ConfusionMatrix}, the back-triple gesture had the highest accuracy, which reached over 99.1\%. Among all different gestures, the left-hold-gesture had the lowest accuracy of 81.9\%. By analyzing the false positive from the confusion matrix, we found that the right-hold-gesture is most likely to be falsely recognized as back-hold-gesture (9.1\%). Overall, confusion was more prominent among similar gestures, such as single and holding gestures (holding gesture is a single tap gesture with a delayed release).

\subsubsection{Comparison between Single and Double Earpieces}

To understand whether having both earpieces are necessary and which gestures are less prone to errors by only having sensors on one side, we further analyzed the accuracy with Left-only earpiece or right-only earpiece. We used the segmentation data saved in the training session and two testing sessions and followed the same data processing pipeline as both earpieces. We found that the average accuracy dropped 18.4\% to 72.4\% for only using the left earpiece, and it dropped 16.0\% to 74.8\% for the right one (Fig. \ref{fig:accuraciesFirsttwotest}). 

\begin{figure}[h]
\centering
\includegraphics[width=1\columnwidth]{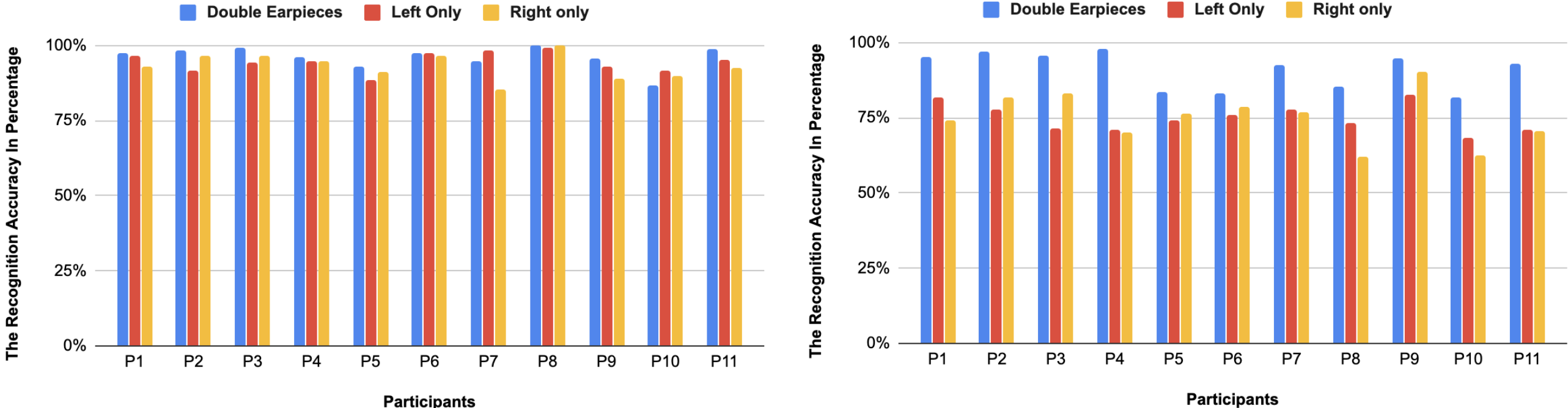}
\caption{a) The accuracy of different channels for `manner' gestures b) The accuracy of different channels for four teeth-tap positions}
\label{fig:AccuraciesForDifferentChannels}
\Description{This figure contains two sub-figures. Both of them are bar charts. X-axis is each participant and Y-axis is the recognition accuracy in percentage. For each participant, both sub-figures shows three bars in blue (double earpieces), red (left only), and yellow (right only).}
\end{figure}

In our gesture set, there are three ``manners'' of contact: double-tap, single-tap, and hold. To understand whether having a single earpiece affects the accuracy, we relabeled the data by only dividing it into three groups: double-tap, single-tap, and hold. From the results (Fig. \ref{fig:AccuraciesForDifferentChannels} a), we first found that the average accuracy across all eleven participants in the testing sessions reached 96.1\% to recognize these three different gesture groups by using both earpieces. By only having single side earpiece, we found the average accuracy only decreased by 1.6\% for the left earpiece and 2.9\% for the right earpiece, respectively. Therefore, we can conclude that using a single side earpiece could reach relatively similar accuracy as double-side earpieces to classify among single-tap gestures, double-tap gestures, and hold gestures. 

To understand the effect of earpiece positions on different teeth-contact areas, we relabeled the data as front-teeth-tap, back-teeth-tap, right-teeth-tap, and left-teeth-tap. We found that the average accuracy in classifying the four kinds of gesture groups with both-side earpieces stayed around 90.9\% (Fig. \ref{fig:AccuraciesForDifferentChannels} b). However, the accuracy decreased dramatically to 74.9\% with the left only earpiece and 75.1\% with right only earpiece, respectively. From the results, we found that the accuracy with both-side earpieces are vital to recognize teeth gestures are different positions.

To further explore the accuracy correlation between the position of earpiece placement and the teeth-tap position, we first conducted a comparative analysis of the accuracy of left-teeth-tap gestures with the left only earpiece and right only earpiece. For the three left-teeth-tap gestures (i.e., `left-single-tap,' `left-double-tap,' and `left-hold'), the average accuracy with left only earpiece (93.2\%) outperformed 3.3\% than the right earpiece (89.9\%). On the other side, we also analyzed the same results for right-teeth-tap gestures (i.e., `right-single-tap,' `right-double-tap,' and `right-hold'). We found that the average accuracy with a left only earpiece (88.4\%) was 5.8\% less than the right one (94.5\%). Therefore, the accuracy is higher for a single earpiece the recognize the teeth gestures that reside on the same side as the earpiece.

\subsubsection{Remounting Effects and Subjective Feedback}

In our study, we conducted a remounting session to understand whether taking the earpieces off and putting them back would affect the accuracy. Overall, we found the average accuracy of recognizing 13 gestures across 11 participants reached 85.3\%, which dropped 5.5\% from the testing sessions. Therefore, we agree that having the participant remount the sensor by themselves may affect the recognition performance. By analyzing the accuracy changes across different participants, we found that P5 (-12.1\%), P3 (-11.5\%), and P9 (-10.9\%) had the accuracy dropped over 10\% in the remounting session. After finishing the study, P5 mentioned his experiences of the remounting session and concerns on making sure the system stays at the relatively same position every time:

\begin{quote}
    ``...To be honest, I forgot where the previous position was after I took it off and trying to put it back. Therefore, I would recommend the researchers to design the artifact that fit on a fixed position on my ears, such as using my ears' shape and force to keep the sensor at the same position, just like the sporting earphones, they always fix at the same place when I use them...''
\end{quote}

We further analyzed the results to uncover how does remounting affect the performance of different gesture sets. For `manner' gestures, we found that the performance only dropped 3\% from 96.1\% to 93.1\% after the participants remounted our prototype. Therefore, we revealed that `manner' gestures are less influenced by remounting the devices.

\TrackChange{After about an hour of the study, our participants completed about 325 gestures, no one reported that they were fatigued. P4 specifically mentioned the benefit of teeth gestures on privacy and `faster response':}

\begin{quote}
    \TrackChange{``...I think the key benefit of having teeth gestures as another input modality is you can make instant responses without even take out the smart devices. Such as playing or pausing music, taking a phone call, I could simply use teeth taps to interact with my smart devices, especially I want this to be applied to my Airpods. Another benefit is that nobody else knows what I did, this will be very useful if I want to reject a call in a meeting...''}
\end{quote}

\section{Discussion}
From our study and findings, we uncovered the feasibility of leveraging IMU sensors on earpieces to track teeth gestures with an accuracy of 90.9\% on average. We also introduced what gesture sets were more error-prone to a single earpiece and whether the subjective feedback and design implications for remounting the device. In this section, we further discuss activation gestures for in-the-wild scenarios, and the opportunities and challenges of deploying TeethTap in real-world future applications. 

\subsection{In-the-wild Scenario}

From the findings, we found that TeethTap successfully recognized 99.1\% for the back-triple gesture. We then conducted a short evaluation to explore whether the back-triple gesture could function as an activation gesture, such as "Hey Siri," to reduce the concerns from false positives. For the "in-the-wild" evaluation, we evaluated how well the activation gesture works while conducting different daily activities. For the same participant group, they were instructed to conduct the following activities in sequence: talking with the researcher, writing on a paper while talking, walking or running around the lab, and eating or drinking. At ten randomized intervals during this process, the researcher asked the participant to perform an activation gesture. Throughout this process, TeethTap was running in real-time on a laptop to detect activation gestures (binary classification). If the performed activation gesture was not detected, we counted the attempt as a false-negative error. If the participant did not perform a gesture and the system detected an activation gesture, we counted this as a false-positive error. The recognition model was built using the five activation gesture instances collected in the previous training sessions. 

Eleven participants tested the activation gesture while performing various activities over a total span of 71 minutes and 33 seconds. Among all Eleven participants, zero false-positive errors were triggered. However, we detected 23 false-negative errors from the 133 gestures. One thing worth mentioning here is that the training data for the activation gesture (back-triple gesture) was collected in the training session while the participants were sitting still in a chair. The added motion introduced from the prescribed activities likely influenced recognition performance. However, we intentionally designed the system to avoid false-positive errors while being more tolerant of false-negative errors, since false-positive errors arguably interfere more with performing daily activities. Therefore, future research could leverage a similar approach to generate an activation gesture to prevent false positives. 

\subsection{Applications and Gesture Sets}
TeethTap offers up to 13 discrete teeth input gestures with an average accuracy rate of 90.9\%. Our participants in our study showed strong interests in embedding teeth gestures to control their smart devices. 

However, to interact with most applications, we may not need to recognize all 13 input gestures at the same time. In other words, a subset of the 13 gestures may be enough for many applications, enabling an even higher accuracy rate. In this section, we discuss potential applications of TeethTap and map possible gesture subsets to each application. 

\subsubsection{Navigating through audio or video content}
The task of navigating through audio or video content could call for the following five gestures: back single (pause/play), left single (previous track), right single (next track), left double (rewind), and right double (fast-forward). The accuracy of recognizing these five gestures is 92.3\% using training and testing data from the user study.

\subsubsection{Volume control}
The holding gesture is designed to provide continuous input, such as changing the volume. A user could simply hold down a gesture to raise or lower volume and release the gesture when the volume has reached the desired level. In this application, only two gestures would be needed: left hold (turn down the volume) and right hold (turn up the volume). The accuracy of recognizing these two gestures is 93.2\% using training and testing data from the user study.

\subsubsection{Operating phone call or videochat}
Phone calls often come at socially inappropriate times. TeethTap could provide discreet gestures that are eyes-free and hands-free to operate a call with two gestures: back double (accept the call) and back single (reject call/hand up). The accuracy of recognizing these two gestures is 98.6\% using training and testing data from the user study. Even in the remounting session, we found that TeethTap could still successfully recognize these two gestures at an accuracy of 94.6\%.

\subsection{Integrating TeethTap to existing head-worn devices}
The current form factor is an independent earpiece, as shown in Figure \ref{fig:earpiece1}. However, we envision TeethTap could be easily adopted into the form factor of existing earphones, headphones, VR headsets or Glass frame technologies. The key integration step is to attach IMUs and contact microphones behind the ear. The form factor could be an extended piece attaching to a Glass frame. Sensors could also be embedded in headphones, following the curvature of the headphone ear-pad around the back of the ear. We believe that integrating such a change would require only hardware alterations, with no changes in the algorithm being necessary. 

\subsection{Improving the performance of session-independent models}
The goal of TeethTap is to provide a user-dependent, but session-independent technology to recognize discrete teeth gestures. In other words, the user needs to provide a few training samples ( e.g. five instances per gesture) when they use TeethTap for the first time. However, they should not have to recollect training data every time they wear the device. The testing results from the fifth session showed that after taking off the device and putting it back on again, the recognition accuracy of TeethTap decreased to around 85.3\%. Apparently, there is room for improvement in the performance.

There are several potential solutions that can help improve TeethTap's performance as a session-independent input technology. Firstly, we can improve the design of the form factor, by improving its precision in applying consistent pressure to the same areas of the body every time it is put on. After all, form factor displacement between sessions is the primary reason for this performance decrease across sessions. Secondly, we can further process sensor data (e.g. normalization) to account for deviations in form factor positioning. Lastly, we could also consider utilizing acoustic sensor data in the gesture classification process (not just the segmentation process), as wearing a position would likely have less of an effect on the acoustic sensor. For instance, we can calculate energy differences between the left and right acoustic sensors to reliably dictate whether an incoming gesture comes from the left side or right side of the mouth.  

\section{Limitation and Future Work}
TeethTap demonstrates the proof-of-concept for detecting a rich set of discrete teeth gestures using an earpiece. However, we do find several limitations and future work from our user evaluation and prototype design.
In the current user study, we did not evaluate TeethTap's performance in recognizing 13 gestures when the user is in motion (e.g. walking, running), which can be limiting in the context of daily life. In the future, we plan to further optimize our system to be functional while the user is moving. There are a few solutions we plan to explore to achieve this feat: 1) we plan to build two separate models---one for static posture and the other for motion---allowing our system to toggle between modes depending on context; 2) we plan on collecting a larger set of training samples and using more advanced machine learning techniques. Furthermore, our participants were from 21 to 34, which lead to being unknown about how well do aging population perform in our study by using our system. In future work, we will further conduct a study with older adults and also discover how well does TeethTap help people with motor impairments who have problems using their smart devices to provide input commands. Although we claimed the existing limitations of our current work, we do believe the current approach has proved the feasibility of leveraging motion tracking on earpieces and combined with noise-filtering from acoustic sensing to recognize different teeth gestures.

\section{Conclusion}
In this paper, we present TeethTap, a wearable technology that can recognize up to 13 discrete teeth gestures. It uses an earpiece which attaches an IMU sensor and a contact microphone behind both the left and right ears. A KNN-based (with the distance measurement of DTW) algorithm is developed for gesture recognition. A user study with 11 participants shows that it can recognize 13 gestures with an accuracy of 90.9\%. We also uncovered the importance of having both-side earpiece available when recognizing position-based gestures comparing with the left-only or right-only earpiece. We further showed the sufficiency of only using a single earpiece to leverage motion sensing to recognize `manner' based gestures. In the discussion, we introduced the approach of reducing false positives through an in-the-wild evaluation with an activation gesture. We also discussed the opportunity and challenges of widely deploying TeethTap on real-world devices in the future. We believe that by fusing motion and acoustic sensing into a minimalist earpiece, TeethTap offers a promising set of novel eyes-free interaction gestures for future applications.


\begin{acks}
This work is supported by Information Science Department at Cornell University. We thank participants for participating the study, reviewers for their thoughtful feedback, and lab members in Cornell SciFi lab for their early feedback on the paper and system design. 

\end{acks}

\bibliographystyle{ACM-Reference-Format}
\bibliography{main}

\end{document}